\documentclass[%
 aip,
cp,  
 amsmath,amssymb,
 reprint,%
]{revtex4-2}

\usepackage{graphicx}
\usepackage{dcolumn}
\usepackage{bm}

\usepackage[utf8]{inputenc}
\usepackage[T1]{fontenc}
\usepackage{mathptmx}

\begin{document}

\title{Quantification of the hadronic CP violation contribution to the atomic EDMs}

\author{Nodoka Yamanaka} 
 \email[Corresponding author: ]{nyamanaka@kmi.nagoya-u.ac.jp}
\affiliation{
Kobayashi-Maskawa Institute for the Origin of Particles and the Universe, Nagoya University, Furocho, Chikusa, Aichi 464-8602, Japan
}

\date{\today} 

\begin{abstract}
CP violating interactions are required to realize the matter abundance of our Universe. 
It is however known that the standard model of particle physics does not contain sufficient CP violation.
The electric dipole moment (EDM) of atomic systems is a very sensitive experimental probe of CP violation beyond the standard model, and it is very actively studied in experiments.
The atomic EDM has a significant sensitivity to hadronic CP violation, but its quantification has for long been obstructed by the nonperturbative physics of quantum chromodynamics.
Quite recently, the contribution of the Weinberg operator (CP violating three-gluon interaction) to the atomic EDM has been analyzed, and we are almost attaining the quantification era of the CP violating hadronic interaction in the leading order of standard model effective field theory.
In this proceedings contribution, we summarize the current attempt to quantify the hadronic CP violation contribution to the EDM of atoms.
\end{abstract}

\maketitle

\section{\label{sec:intro}Introduction}

The excess of matter over antimatter observed in the Universe is realized if and only if the conditions given by Sakharov \cite{Sakharov:1967dj} are fulfilled.
One of them, the CP violation of the fundamental theory, is known to be in critical deficit.
The particle physics community is currently looking for new physics beyond the standard model (SM) containing large CP violation.
As an observable very sensitive to it, we have the electric dipole moment (EDM) of atoms \cite{Yamanaka:2014mda,Yamanaka:2017mef,Chupp:2017rkp,Blum:2022cie}.
Its SM contribution is known to be very small \cite{Seng:2014lea,Yamanaka:2015ncb,Yamanaka:2016fjj,Lee:2018flm,Yamaguchi:2020eub,Yamaguchi:2020dsy,Ema:2022yra}, which is one of the reasons why it is an attractive experimental probe.

The EDM {\bf d} is the linear response against the external electric field {\bf E}, and its interaction Hamiltonian is 
\begin{equation}
H_{\rm EDM}
=
- {\bf d} \cdot {\bf E}
.
\end{equation}
Its relativistic form is given by the following dimension-5 Lagrangian
\begin{equation}
{\cal L}_{\rm EDM}
=
- \frac{i}{2} d_q \bar q \sigma_{\mu \nu} F^{\mu \nu} \gamma_5 q
.
\end{equation}
This effective interaction may perturbatively be calculated in many new physics models, such as the extended Higgs models, supersymmetry, etc.
At the elementary level, there are also other CP violating operators such as the quark chromo-EDM, the Weinberg operator \cite{Weinberg:1989dx}, and the CP-odd four-quark interactions, defined by
\begin{eqnarray}
{\cal L}_{\rm cEDM}
&=&
- \frac{i}{2} d^c_q \bar q \sigma_{\mu \nu} F_a^{\mu \nu} t_a \gamma_5 q
,
\\
{\cal L}_{w}
&=&
\frac{w}{6} f_{abc} \epsilon_{\alpha \beta \gamma \delta} F_a^{\mu \alpha} F_b^{\beta \gamma} F_{\delta , c}^{\ \ \mu}
,
\label{eq:weinbergop}
\\
{\cal L}_{\rm 4q}
&=&
C_{qq'} \bar q i \gamma_5 q \, \bar q' q'
,
\end{eqnarray}
respectively.
These are the leading CP-odd interactions of the SM effective field theory.

The above CP-odd interactions are known to mix each other when the renormalization scale is changed from the scale beyond the SM, say TeV, to the hadronic one (GeV).
The mixing due to QCD effects is important \cite{Degrassi:2005zd,Hisano:2012cc}, and this forces us to calculate all relevant hadron level matrix elements.
At the hadron level, the leading CP violating processes are the nucleon EDM and the CP-odd pion-nucleon interaction
\begin{equation}
{\cal L}_{\pi NN}
=
\bar g_{\pi NN}^{(0)} \pi_a \bar N \tau_a N
+
\bar g_{\pi NN}^{(1)} \pi_0 \bar N N
+
\bar g_{\pi NN}^{(2)} ( \pi_a \bar N \tau_a N - 3 \pi_0 \bar N \tau_3 N)
,
\label{eq:piNN}
\end{equation}
which contributes to the nuclear EDM and to the nuclear Schiff moment \cite{Schiff:1963zz}, the CP-odd nuclear moment contributing to the atomic EDM, via the one-pion exchange CP-odd nuclear force.

In this proceedings contribution, we review the current status of the theoretical calculations of the hadronic matrix elements of CP-odd quark-gluon operators contributing to the nucleon EDM and to Eq. (\ref{eq:piNN}).
Once we obtain them, we can proceed to the evaluation of the nuclear and atomic EDMs.
The nuclear and atomic level calculations have uncertainties up to 30\% for many interesting systems \cite{Yamanaka:2017mef,Chupp:2017rkp,Yamanaka:2015qfa,Yamanaka:2016itb,Yamanaka:2016umw,Yamanaka:2019vec,Froese:2021civ,Yanase:2020agg,Yanase:2020oos,Yanase:2022atk,Dobaczewski:2005hz,Sakurai:2019vjs,Hubert:2022pnl,Prasannaa:2020cjx}, but the hadron level ones have much larger errorbars.
In the next section, we see one by one the currently known values of the hadron level CP-odd couplings generated by the quark EDM, the theta-term, the CP-odd four-quark interaction, the Weinberg operator, and the quark chromo-EDM, as well as the methods used to derive them.
We also present the current status of the pion-nucleon sigma term which is very important in the quantification of the above couplings.
The final section is devoted to the summary.

\section{\label{sec:hadmatele}Review of CP-odd hadron matrix elements}

\subsection{\label{sec:quark_edm}Nucleon EDM from quark EDM}

The quark EDM mainly contributes to the nucleon EDM.
The coefficients of proportionality relating them are called nucleon tensor charges, and they have been extensively calculated in lattice QCD \cite{FlavourLatticeAveragingGroupFLAG:2021npn,Yamanaka:2018uud}.
The averaged result for the up and down quark contributions to the neutron EDM is
\begin{equation}
d_n 
\approx
0.8 d_d 
-0.2 d_u
,
\end{equation}
with about 10\% of error (for the proton EDM, just interchange up and down quarks, from isospin symmetry).
This is actually the most successful case of lattice QCD calculations in the context of the EDM.

\subsection{\label{sec:theta_term}Theta-term contribution}

The theta-term mainly contributes to the nucleon EDM and to the CP-odd nuclear force via the isoscalar type CP-odd pion-nucleon interaction [see the term with $\bar g_{\pi NN}^{(0)}$ of Eq. (\ref{eq:piNN})].
In the leading order of chiral perturbation, the generated neutron EDM is \cite{Crewther:1979pi}
\begin{equation}
d_n 
\approx
\frac{e g_A m_N  \bar g_{\pi NN}^{(0)} }{4 \pi^2 f_\pi} \ln \frac{m_N}{m_\pi}
\approx
-0.0025 \theta e \, {\rm fm}
,
\end{equation}
where the isoscalar CP-odd coupling depends on $\theta$ as $\bar g_{\pi NN}^{(0)} = \frac{m_u m_d \theta}{f_\pi (m_u +m_d)} \langle N| \bar uu - \bar d d|N \rangle$.
This value is consistent with the available lattice QCD result $d_n = -0.00152 (71) \theta e$ fm \cite{Dragos:2019oxn}, although it has a large uncertainty.

\subsection{\label{sec:4-quark}CP-odd four-quark interaction}

The CP-odd four-quark interaction has many possible flavor structures, but it mainly contributes to the atomic EDM via the isovector CP-odd pion-nucleon interaction.
The most quantifiable way to evaluate its coupling constant to date is the factorization \cite{An:2009zh}.
The factorization with the vacuum saturation approximation works as
\begin{equation}
\langle \pi_0 N | \bar q i \gamma_5 q \, \bar q' q' | N \rangle 
\approx
\langle \pi_0 | \bar q i \gamma_5 q | 0 \rangle \langle N |\bar q' q' | N \rangle 
.
\end{equation}
We then obtain a product of two known matrix elements.
This approximation is good in the large $N_c$ limit if the quarks $q$ and $q'$ have different flavors.
In the real QCD case, the contribution from higher order terms, which constitutes the uncertainty, is estimated to be of $O(60\% )$.

\subsection{\label{sec:weinbergop}Weinberg operator}

The Weinberg operator (\ref{eq:weinbergop}) is a purely gluonic operator, so the chiral perturbation cannot be used.
The leading contribution is currently known to be given by the nucleon EDM generated by the chiral rotation of the nucleon anomalous magnetic moment, and the rotation angle has been evaluated using QCD sum rules \cite{Demir:2002gg,Haisch:2019bml}.
The determination of the systematics due to remaining CP violating contributions has also been achieved quite recently \cite{Yamanaka:2020kjo,Osamura:2022rak,Yamanaka:2022qlu}.
The atomic EDM receives the leading contribution from the Weinberg operator via the neutron EDM
\begin{equation}
d_n 
=
(20 \pm 12) w\, e \, {\rm MeV}
.
\end{equation}
However, we have to note that the pion-exchange contribution might not be negligible for heavy atoms \cite{Osamura:2022rak}.

\subsection{\label{sec:chromo-EDM}Chromo-EDM}

The quark chromo-EDM is generated in many known candidates of new physics, so its quantification is of primary importance.
However, its hadron matrix elements are currently having a sizable uncertainty.
The chromo-EDM generates both the isoscalar and isovector CP-odd pion-nucleon interactions, and the former induces a large nucleon EDM, while the latter yields an important contribution to the CP-odd nuclear force.
The main problem of the quantification is that the matrix element $\langle N | \bar \psi \sigma_{\mu \nu} F_a^{\mu \nu} t_a \psi |N\rangle$ is not known accurately.
There are several evaluations using models, but the results have large dispersion, and their orders, and sometimes even their signs, differ \cite{Khatsimovsky:1987fr,Pospelov:2001ys}.
Unfortunately, we can only say that the error bar of the chromo-EDM contribution to the isovector coupling is $O(100\%)$, centered at \cite{Khatsimovsky:1987fr}
\begin{equation}
\bar g_{\pi NN}^{(1)}
\approx
-1.5 \times 10^{-11} \frac{d_u^c-d_d^c}{10^{-26} {\rm cm}}
.
\end{equation}

\subsection{\label{sec:sigma_term}Pion-nucleon sigma term}

The pion-nucleon sigma term $\sigma_{\pi N} \equiv \frac{m_u + m_d}{\langle N | \bar uu +\bar dd |N\rangle}$ is a well-known nucleon matrix element for being an important input in chiral perturbation theory.
It is also almost maximally relevant in the calculations of the leading order contributions to the nucleon EDM and to the CP-odd nuclear force.
It is also used in the calculation of the atomic EDM induced by the CP-odd electron-nucleon interaction \cite{Yanase:2018qqq}, so accurately knowing $\sigma_{\pi N}$ is crucial for the analysis of CP violation beyond the SM.

\begin{figure}
\includegraphics[width=7.5 cm]{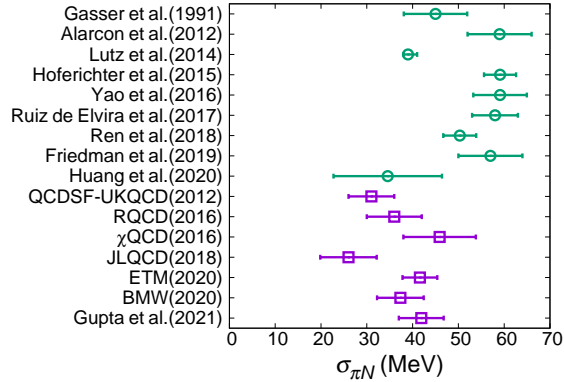}
\caption{\label{fig:sigma_term}
Comparison of known results of the evaluations of the pion-nucleon sigma term.
The data points of phenomenological extractions and lattice QCD calculations are drawn in green and purple, respectively.
This figure was adapted from Ref. \cite{Yamanaka:2018uud}, with data taken from Refs. \cite{FlavourLatticeAveragingGroupFLAG:2021npn,Gupta:2021ahb}.
}
\end{figure}

However, it is currently known that the results of the evaluations of $\sigma_{\pi N}$ are not agreeing between phenomenological extractions ($\approx 60$ MeV) and lattice QCD ($\approx 30$ MeV) \cite{FlavourLatticeAveragingGroupFLAG:2021npn,Yamanaka:2018uud,Gupta:2021ahb}.
We plot in Fig. \ref{fig:sigma_term} the current situation of $\sigma_{\pi N}$.
The reason of this discrepancy is not known, and this is introducing a sizable systematics, for which a resolution is needed in the future.

\section{Conclusion}

In this proceedings contribution, we reviewed the current status of the quantification of the hadronic matrix elements needed in the evaluation of the EDM of atoms.
We are almost reaching the quantification era of the hadronic CP violation as regards the EDM thanks to the progress in the analysis of the Weinberg operator contribution.
Now the most important work to be done is the improvement of the quark chromo-EDM matrix elements.

\begin{acknowledgments}
NY was supported by Daiko Foundation.
\end{acknowledgments}

\nocite{*}
\bibliography{yamanaka}

\providecommand{\href}[2]{#2}\begingroup\raggedright\begin{thebibliography}{10}

\bibitem{Sakharov:1967dj}
A.~D.~Sakharov,
Pisma Zh. Eksp. Teor. Fiz. \textbf{5}, 32-35 (1967).

\bibitem{Yamanaka:2014mda}
N.~Yamanaka,
``Analysis of the Electric Dipole Moment in the R-parity Violating Supersymmetric Standard Model,''
Springer, Berlin Germany (2014).

\bibitem{Yamanaka:2017mef}
N.~Yamanaka, B.~K.~Sahoo, N.~Yoshinaga, T.~Sato, K.~Asahi and B.~P.~Das,
Eur. Phys. J. A \textbf{53}, no.3, 54 (2017)
[arXiv:1703.01570 [hep-ph]].

\bibitem{Chupp:2017rkp}
T.~Chupp, P.~Fierlinger, M.~Ramsey-Musolf and J.~Singh,
Rev. Mod. Phys. \textbf{91}, no.1, 015001 (2019)
[arXiv:1710.02504 [physics.atom-ph]].

\bibitem{Blum:2022cie}
T.~Blum \textit{et al.},
[arXiv:2209.08041 [hep-ex]].

\bibitem{Seng:2014lea}
C.~Y.~Seng,
Phys. Rev. C \textbf{91}, no.2, 025502 (2015)
[arXiv:1411.1476 [hep-ph]].

\bibitem{Yamanaka:2015ncb}
N.~Yamanaka and E.~Hiyama,
JHEP \textbf{02}, 067 (2016)
[arXiv:1512.03013 [hep-ph]].

\bibitem{Yamanaka:2016fjj}
N.~Yamanaka and E.~Hiyama,
Nucl. Phys. A \textbf{963}, 33-51 (2017)
[arXiv:1605.00161 [nucl-th]].

\bibitem{Lee:2018flm}
J.~Lee, N.~Yamanaka and E.~Hiyama,
Phys. Rev. C \textbf{99}, no.5, 055503 (2019)
[arXiv:1811.00329 [nucl-th]].

\bibitem{Yamaguchi:2020eub}
Y.~Yamaguchi and N.~Yamanaka,
Phys. Rev. Lett. \textbf{125}, 241802 (2020)
[arXiv:2003.08195 [hep-ph]].

\bibitem{Yamaguchi:2020dsy}
Y.~Yamaguchi and N.~Yamanaka,
Phys. Rev. D \textbf{103}, no.1, 013001 (2021)
[arXiv:2006.00281 [hep-ph]].

\bibitem{Ema:2022yra}
Y.~Ema, T.~Gao and M.~Pospelov,
[arXiv:2202.10524 [hep-ph]].

\bibitem{Weinberg:1989dx}
S.~Weinberg,
Phys. Rev. Lett. \textbf{63}, 2333 (1989).

\bibitem{Degrassi:2005zd}
G.~Degrassi, E.~Franco, S.~Marchetti and L.~Silvestrini,
JHEP \textbf{11}, 044 (2005)
[arXiv:hep-ph/0510137 [hep-ph]].

\bibitem{Hisano:2012cc}
J.~Hisano, K.~Tsumura and M.~J.~S.~Yang,
Phys. Lett. B \textbf{713}, 473-480 (2012)
[arXiv:1205.2212 [hep-ph]].

\bibitem{Schiff:1963zz}
L.~I.~Schiff,
Phys. Rev. \textbf{132}, 2194-2200 (1963).

\bibitem{Yamanaka:2015qfa}
N.~Yamanaka and E.~Hiyama,
Phys. Rev. C \textbf{91}, no.5, 054005 (2015)
[arXiv:1503.04446 [nucl-th]].

\bibitem{Yamanaka:2016itb}
N.~Yamanaka, T.~Yamada, E.~Hiyama and Y.~Funaki,
Phys. Rev. C \textbf{95}, no.6, 065503 (2017)
[arXiv:1603.03136 [nucl-th]].

\bibitem{Yamanaka:2016umw}
N.~Yamanaka,
Int. J. Mod. Phys. E \textbf{26}, no.4, 1730002 (2017)
[arXiv:1609.04759 [nucl-th]].

\bibitem{Yamanaka:2019vec}
N.~Yamanaka, T.~Yamada and Y.~Funaki,
Phys. Rev. C \textbf{100}, no.5, 055501 (2019)
[arXiv:1907.08091 [nucl-th]].

\bibitem{Froese:2021civ}
P.~Froese and P.~Navratil,
Phys. Rev. C \textbf{104}, no.2, 025502 (2021)
[arXiv:2103.06365 [nucl-th]].

\bibitem{Yanase:2020agg}
K.~Yanase and N.~Shimizu,
Phys. Rev. C \textbf{102}, no.6, 065502 (2020)
[arXiv:2006.15142 [nucl-th]].

\bibitem{Yanase:2020oos}
K.~Yanase,
Phys. Rev. C \textbf{103}, no.3, 035501 (2021)
[arXiv:2008.03678 [nucl-th]].

\bibitem{Yanase:2022atk}
K.~Yanase, N.~Shimizu, K.~Higashiyama and N.~Yoshinaga,
[arXiv:2210.08498 [nucl-th]].

\bibitem{Dobaczewski:2005hz}
J.~Dobaczewski and J.~Engel,
Phys. Rev. Lett. \textbf{94}, 232502 (2005)
[arXiv:nucl-th/0503057 [nucl-th]].

\bibitem{Sakurai:2019vjs}
A.~Sakurai, B.~K.~Sahoo, K.~Asahi and B.~P.~Das,
Phys. Rev. A \textbf{100}, no.2, 020502 (2019)
[arXiv:1908.04151 [physics.atom-ph]].

\bibitem{Hubert:2022pnl}
M.~Hubert and T.~Fleig,
Phys. Rev. A \textbf{106}, no.2, 022817 (2022)
[arXiv:2203.04618 [physics.atom-ph]].

\bibitem{Prasannaa:2020cjx}
V.~S.~Prasannaa, R.~Mitra and B.~K.~Sahoo,
J. Phys. B \textbf{53}, no.19, 195004 (2020).

\bibitem{FlavourLatticeAveragingGroupFLAG:2021npn}
Y.~Aoki \textit{et al.} [Flavour Lattice Averaging Group (FLAG)],
Eur. Phys. J. C \textbf{82}, no.10, 869 (2022)
[arXiv:2111.09849 [hep-lat]].

\bibitem{Yamanaka:2018uud}
N.~Yamanaka \textit{et al.} [JLQCD],
Phys. Rev. D \textbf{98}, no.5, 054516 (2018)
[arXiv:1805.10507 [hep-lat]].

\bibitem{Crewther:1979pi}
R.~J.~Crewther, P.~Di Vecchia, G.~Veneziano and E.~Witten,
Phys. Lett. B \textbf{88}, 123 (1979)
[erratum: Phys. Lett. B \textbf{91}, 487 (1980)].

\bibitem{Dragos:2019oxn}
J.~Dragos, T.~Luu, A.~Shindler, J.~de Vries and A.~Yousif,
Phys. Rev. C \textbf{103}, no.1, 015202 (2021)
[arXiv:1902.03254 [hep-lat]].

\bibitem{An:2009zh}
H.~An, X.~Ji and F.~Xu,
JHEP \textbf{02}, 043 (2010)
doi:10.1007/JHEP02(2010)043
[arXiv:0908.2420 [hep-ph]].

\bibitem{Demir:2002gg}
D.~A.~Demir, M.~Pospelov and A.~Ritz,
Phys. Rev. D \textbf{67}, 015007 (2003)
[arXiv:hep-ph/0208257 [hep-ph]].

\bibitem{Haisch:2019bml}
U.~Haisch and A.~Hala,
JHEP \textbf{11}, 154 (2019)
[arXiv:1909.08955 [hep-ph]].

\bibitem{Yamanaka:2020kjo}
N.~Yamanaka and E.~Hiyama,
Phys. Rev. D \textbf{103}, no.3, 035023 (2021)
[arXiv:2011.02531 [hep-ph]].

\bibitem{Osamura:2022rak}
N.~Osamura, P.~Gubler and N.~Yamanaka,
JHEP \textbf{06}, 072 (2022)
[arXiv:2203.06878 [hep-ph]].

\bibitem{Yamanaka:2022qlu}
N.~Yamanaka and M.~Oka,
Phys. Rev. D \textbf{106}, no.7, 075021 (2022)
[arXiv:2208.03920 [nucl-th]].

\bibitem{Khatsimovsky:1987fr}
V.~M.~Khatsimovsky, I.~B.~Khriplovich and A.~S.~Yelkhovsky,
Annals Phys. \textbf{186}, 1-14 (1988).

\bibitem{Pospelov:2001ys}
M.~Pospelov,
Phys. Lett. B \textbf{530}, 123-128 (2002)
[arXiv:hep-ph/0109044 [hep-ph]].

\bibitem{Yanase:2018qqq}
K.~Yanase, N.~Yoshinaga, K.~Higashiyama and N.~Yamanaka,
Phys. Rev. D \textbf{99}, no.7, 075021 (2019)
[arXiv:1805.00419 [nucl-th]].

\bibitem{Gupta:2021ahb}
R.~Gupta \textit{et al.},
Phys. Rev. Lett. \textbf{127}, no.24, 24 (2021)
[arXiv:2105.12095 [hep-lat]].

\end{thebibliography}\endgroup

\end{document}